# Neural nanophotonic object detector with ultra-wide field-of-view

## Author Information

Ji Chen[1,2,4]*, Yue Wu[1,4], Muyang Li[1,4], Zhongyi Yuan[1,4], Zi-Wen Zhou[1], Cheng-Yao Hao[1], Bingcheng Zhu[1,2], Yin Wang[2], Jitao Ji[3], Chunyu Huang[3], Haobai Li[1], Yanxiang Zhang[1], Kai Qiu[3], Shining Zhu[3], Tao Li[3]* & Zaichen Zhang[1,2]*

1. School of Information Science and Engineering, National Mobile Communications Research Laboratory, Frontiers Science Center for Mobile Information Communication and Security, **Quantum Information Research Center,** Southeast University, Nanjing 210096, China.

2. Purple Mountain Laboratories, Nanjing 211111, China.

3. National Laboratory of Solid State Microstructures, College of Engineering and Applied Science, School of Physics, Nanjing University, Nanjing, 210023, China.

4. These authors contributed equally: Ji Chen, Yue Wu, Muyang Li, Zhongyi Yuan.

## Contributions

J. C. and Z. Z. developed the idea. J. C., Y. Wu and Z.-W. Z proposed the design and performed the numerical simulation. Y. Wu, Z.-W. Z., M. L. and Z. Y. conducted the optical testing experiments with the help from J. C. and Y. Wang.. M. L., C.-Y. H. and Y. Z. conducted the neural network calculation with the assistance from J. C. and H. L.. Y. Wu., C. H. and J. J. fabricated and characterized the samples with the help from K. Q.. Z. Y. and M. L. conducted the device integration under the guidance of B. Z.. J. C., T. L. and Z. Z. supervised the project. J.C. analyzed the results and wrote the manuscript under the guidance of S. Z.. All authors contributed to the discussion.

## Corresponding authors

Correspondence and requests for materials should be addressed to Ji Chen, jichen@seu.edu.cn, Tao Li, taoli@nju.edu.cn, and Zaichen Zhang, zczhang@seu.edu.cn.

## Abstract

Intelligent object detection, which extracts crucial information like targets categories and locations, plays a vital role in emerging technologies including autonomous driving, the Internet of Things, and next-generation mobile communication systems. With the advancement of intelligent object detectors towards higher integration and miniaturization, their portability and adaptability to a broader range of scenarios have been significantly enhanced. However, this progress comes at the cost of reduced detection quality and narrower field-of-view, which severely impacts overall performances. Here we present a neural nanophotonic object detector based on a metalens array, capable of delivering high-quality imaging with an ultra-wide field-of-view of 135°. The combined neural network not only further improves the imaging quality, but also enables the detector to achieve high-precision target recognition and localization. Moreover, we integrated the neural nanophotonic object detector into a miniature unmanned aerial vehicle to enable wide-angle imaging and intelligent recognition of various real-world dynamic objects, demonstrating the high mobility and flexibility of our neural nanophotonic object detector. Our study presents a systematic framework for advancing revolutionary intelligent detection systems, offering significant potential for a wide range of future applications.

## Introduction

Intelligent object detection refers to automatic identification of objects of interest within a given scene, enabling subsequent information processing or interaction with the identified targets. It has broad applications across various fields, including autonomous driving, security and surveillance, industrial automation, and smart cities[1-5]. Among the available approaches, imaging-based methods powered by computer vision have emerged as a dominant technique due to their numerous advantages, such as high resolution, fast processing speeds, and rich information transmission[6-10].

In scenarios requiring high portability or operations in confined spaces, ultra-compact imaging devices are very crucial. However, size constraints often result in trade-offs in imaging performance, particularly in terms of image quality and field-of-view (FOV), which would significantly affect the accuracy and effective range of subsequent object detection processes.

Metalenses are artificial lenses composed of two-dimensional (2D) subwavelength units[11-16]. Thanks to their ultra-lightweight, ultra-thin and flexible design features, metalenses have been successfully employed to create highly integrated imaging devices with diverse functionalities[17-22]. Certain high-efficiency dielectric metalenses can deliver performance comparable to that of commercial objective lenses[23-27]. However, achieving a large FOV remains a significant challenge for metalenses. Although several wide-FOV metalens designs have been proposed, including two-layer metalens, quadratic phase metalens, and computational thin-plate lenses[28-35], these methods often encounter fabrication or integration difficulties or compromise key imaging performance metrics such as efficiency or resolution. Developing a single-layer metalens with capabilities of both ultra-wide field-of-view (FOV) and high imaging quality is crucial for advancing highly integrated intelligent object detectors, allowing them to detect a wider range of targets effectively in a compact form.

In this work, we propose a neural nanophotonic object detector (NNOD) based on a single layer of a metalens array and a comprehensive neural network, achieving an ultra-wide FOV of 135° (**Fig. 1a**). Three metalenses of the metalens array are meticulously designed to provide clear imaging for different angular ranges (**Fig. 1b**). The sub-images produced by these metalenses collectively form a complete wide-angle image through processes like distortion correction and stitching. A comprehensive neural network is then employed to further enhance the captured image's quality and intelligently identifying target objects (**Fig. 1c**). These advancements enable high object detection precision (>96%), large detection distance (>10m) and precise angular localization (<0.1°). Additionally, we presented an integration solution of the NNOD with a miniature unmanned aerial vehicle (MAV) featuring wireless image transmission and high-performance detection of real-world dynamic objects, demonstrating the high portability and integrability.

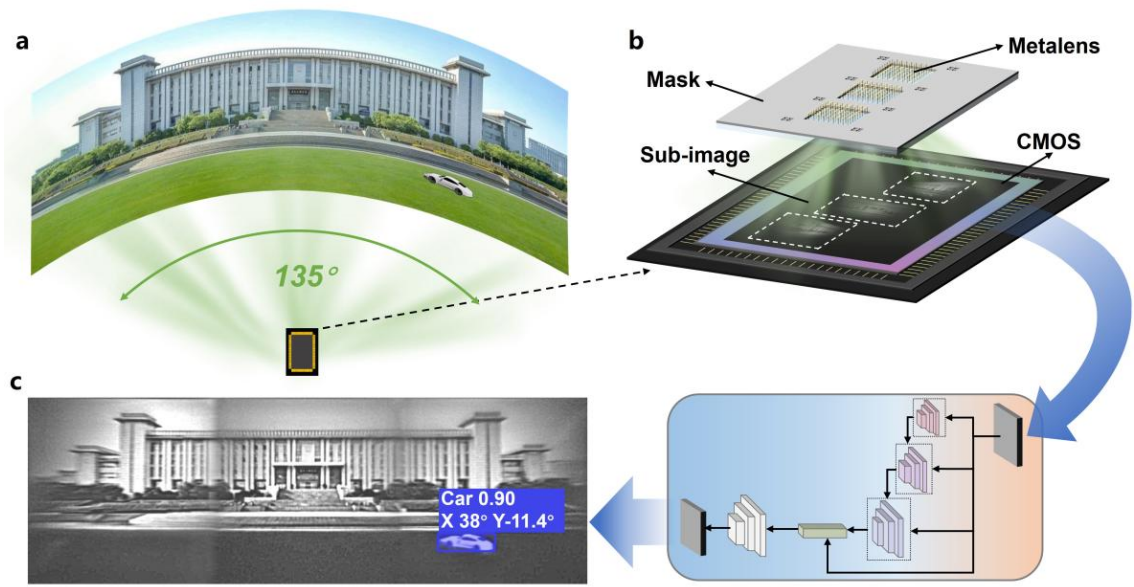

**Fig. 1 | Working principle of the neural nanophotonic object detector (NNOD). a,** The NNOD is capable of imaging a wide scene with a FOV of 135°. **b,** The key hardware components of NNOD are a metalens array and a CMOS sensor. The three metalenses of the metalens array produce clear images of objects from different viewing angles, creating three sub-images. The surrounding mask blocks direct ambient light to enhance image quality. **c,** Three sub-images are then fed into a comprehensive neural network to get a high-quality wide-angle imaging result. The target objects within the wide-angle image are accurately identified and located.

**Design of the wide-angle metalens array**

The core component of the NNOD is a metalens array for wide-angle imaging, comprising three metalenses designed for different viewing angles (one center-FOV metalens and two side-FOV metalenses). However, designing a metalens with an ultra-broad viewing angle is inherently challenging, as it requires the unit cells to adjust phase responses to varying angle of incidence (**Supplementary Information Section 1**). To overcome this challenge, we first employed a multi-configuration ray tracing optimization method to derive the phase profile of the center-FOV metalens (**Fig. 2a** and **Methods**). The basic concept of this approach is to optimize the phase profile by defining specific virtual apertures that direct incident light from various angles to illuminate distinct regions of the metalens. It should be noted that the apertures used in optimization are virtual, rather than physical. When the virtual apertures are removed from the ray tracing model, the actual light modulation of the designed single-layer center-FOV metalens is revealed (**Fig. 2b**). Light passing through the virtual aperture regions is well focused by the metalens, with the focal spots size comparable to that of the Airy disk (**Fig. 2a**). Light outside the virtual aperture regions is evenly distributed across the CMOS sensor, which would be effectively mitigated in subsequent neural network processing. After a comprehensive evaluation, we determined the center-FOV metalens features a diameter of 1 mm, a focal length of 1.6 mm, a working wavelength of 550 nm, and a FOV ranging from -22.5° to 22.5°, respectively. (**Supplementary Information Section 2 and 3**).

The phase profiles of the two side-FOV metalenses require no further optimization, instead they can be obtained by adding a tilt phase $\pm kd\sin\theta_0$ to the center-FOV metalens phase, where $k$ is the wave number, $d$ is the metalens diameter, and $\theta_0$ is the designed tilt angle. Introducing the tilt phase effectively compensates for the additional phase shifts caused by oblique incidence, thereby enabling a precise adjustment of the metalens imaging angular range. Here $\theta_0$ is set as 45° to make the two side-FOV be -67.5° to -22.5° and 22.5° to 67.5°, to symmetrically offset the center-FOV. Ray tracing analysis of the two side-FOV (**Fig. 2b**) show that major rays are focused to high-quality spots like the center-FOV metalens, ensuring the wide-angle imaging capability of the entire metalens array.

**Fabrication and optical characterizations of the wide-angle triplet metalens array**

Three metalenses were fabricated on a silica ($SiO_2$) substrate, arranged longitudinally with their imaging FOV oriented laterally to minimize the cross-talk among the sub-images. To further reduce noise from direct light transmission outside the metalenses structure regions, a mask was fabricated (**Fig. 2c**). The metalenses structures were fabricated in silicon nitride ($Si_3N_4$) nano-posts with a height of 1000 nm and varying radius of round cross-sections. Through finite-difference time-domain (FDTD) simulation, eight types of nano-posts with varying cross-sectional radii were identified, which cover the $2\pi$ phase range and all exhibit over 90% transmittance (**Fig. 2d**). The inset displays the top-view of scanning electron microscope (SEM) image of the metalens structures. See **Methods** and **Extended Data Fig. 1** for details of simulation and fabrication.

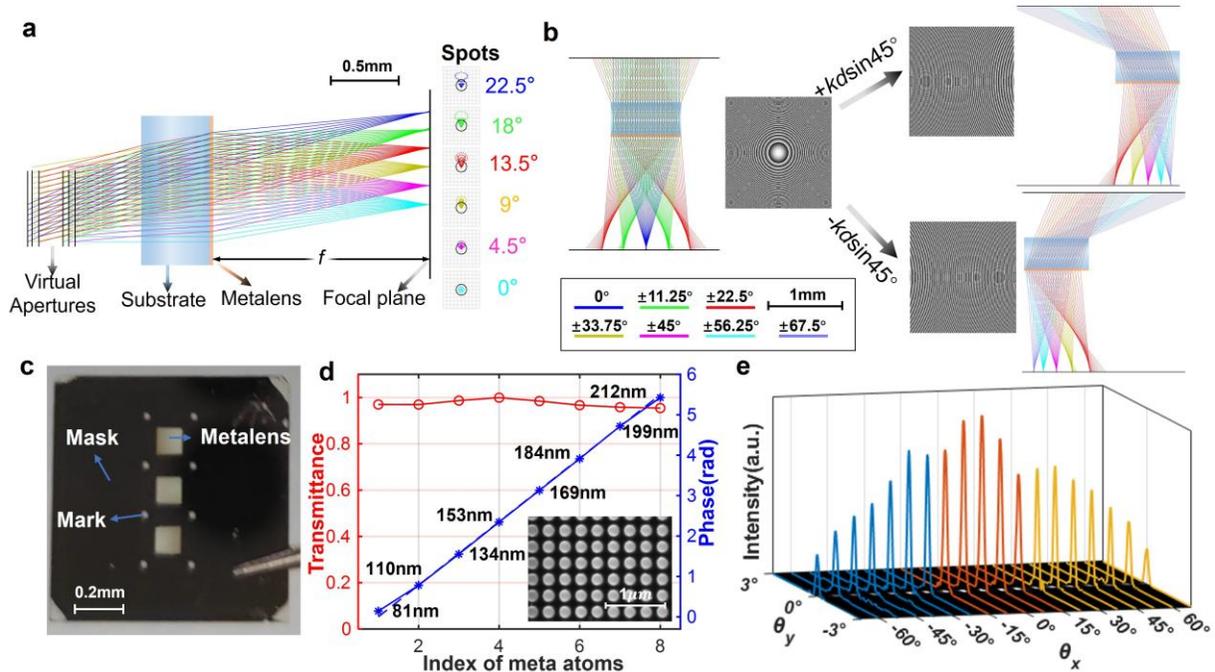

**Fig. 2 | Design and characterization of the wide-angle metalens array. a,** Ray tracing simulation of the center-FOV metalens with virtual apertures. The focal spots of light rays with incident angles of 0°, 4.5°, 9°, 13.5°, 18° and 22.5° are shown at the right side. The black circles indicate the size of the Airy disks corresponding to different incident angles. **b,** Ray tracing simulation of the center-FOV metalens and the two side-FOV metalenses without virtual apertures. The phase profiles of the two side-FOV metalenses are obtained by adding the tilt phase $\pm kd\sin45°$ to the center-FOV metalens phase profile, respectively. **c,** Photograph of the metalens array sample. The three large square regions represent the metalens structure areas, while the small regions around metalenses are markers for alignment in fabrication. The surrounding black areas correspond to the mask. **d,** Phase (blue stars) and transmittance (red circles) of meta-atoms with 8 different structural parameters, simulated by FDTD solutions at the wavelength of 550 nm. The diameters of the nano-posts are marked along the phase distribution line. Inset is the top-view scanning electron

microscopy (SEM) image of part of the nano-structures. **e,** Focal spots and their intensities corresponding to different incident angle. The intensities are normalized to the maximum of the 0° focal spot intensity.

Upon completion of the metalens array fabrication, optical testing was conducted to verify its wide-angle focusing performance (**Methods** and **Extended Data Fig. 2a**). The focal spots of light corresponding to different incident angles were captured and analyzed (**Fig. 2e**). The intensity distribution curves of different focal spots are normalized to the peak value of 0° focal spot intensity. The narrow full width at half maximums (FWHMs) of the curves indicates the focal spots maintain relatively good focusing quality. The modulation transfer functions (MTFs) of the metalens array (**Extended Data Fig. 2b**) were also obtained through the ray tracing analysis. Compared with the MTFs of a traditional hyperbolic phase metalens (**Extended Data Fig. 2c**), the MTFs of our metalens array for different incident angles closely approach the diffraction limit curve, demonstrating significantly improved wide-angle imaging capability.

**Wide-angle imaging of the metalens array**

We initially conducted imaging experiments on the optical platform to determine the parameters for processes of distortion correction and sub-images stitching, and also to acquire the imaging datasets for subsequent neural network training. The wide-angle images were projected onto a display composed of three OLED screens, and then be captured by the metalens array (**Fig. 3a** and **Methods**). Imaging distortion is an inherent challenge in wide-angle imaging, and effectively mitigating it is essential for improving both imaging quality and the accuracy of intelligent object detection and localization. Through the imaging distortion analysis of a 2D dot matrix and a chessboard pattern, a distortion correction method was successfully developed (**Extended Data Fig. 3** and **Methods**). Since imaging distortion is primarily dependent on the imaging device, the distortion mapping is fixed once the metalens array is fabricated. Consequently, the distortion correction method can be applied to any subsequent wide-angle images of the same metalens array.

We first conducted wide-angle imaging of a pure letter pattern of "Southeast University" with a viewing angle of 150°. Due to its simplicity and high contrast, the imaging viewing angle could exceed the designed FOV of 135°. The sub-images captured directly by the CMOS exhibit obvious imaging distortion with distinct regions of clear and blurred imaging (**Fig. 3b**). After distortion correction, the clear imaging region of each sub-image were extracted and stitched together to produce the complete wide-angle image (**Fig. 3c**). For comparison, a traditional hyperbolic phase metalens was also employed to capture the image (**Fig. 3d**). It is evident that our wide-angle metalens array expands the FOV by near three times. Additionally, we imaged a complex scene (the original image shown in **Fig. 1a**), which imposes much higher demands on the imaging capability. Thus, we set the imaging angle to the designed 135°, and followed the distortion correction and stitching processes to obtain the wide-angle imaging result (**Fig. 3e**).

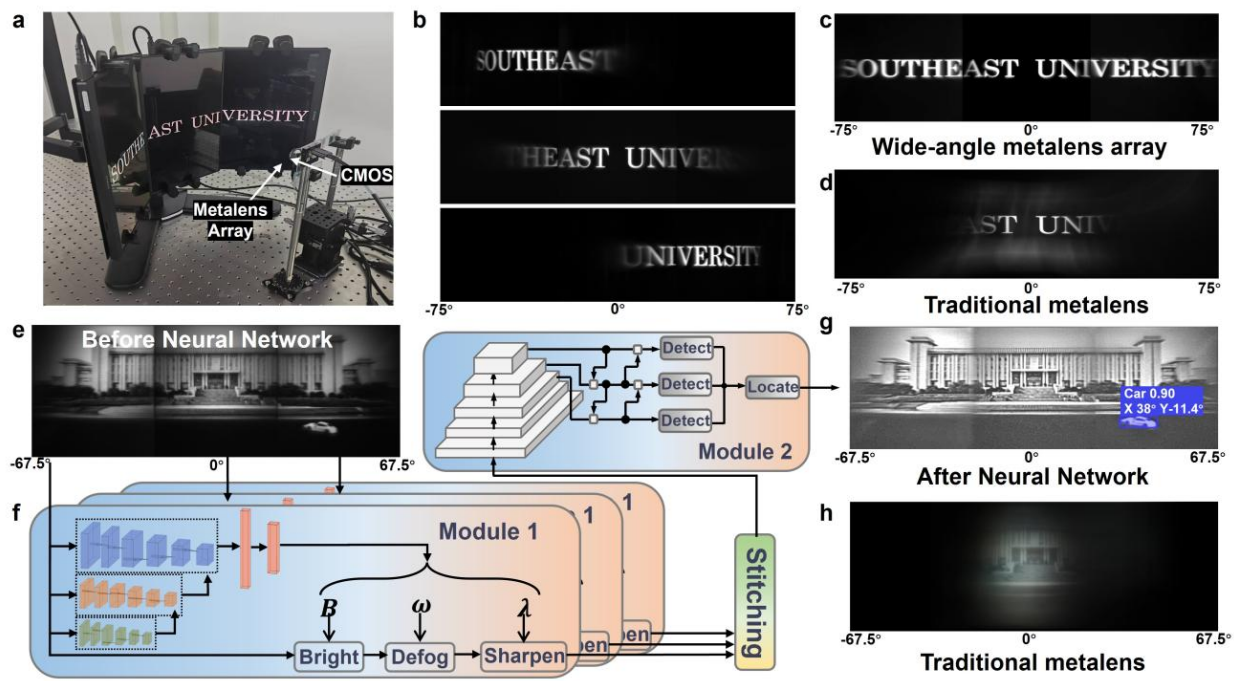

**Fig. 3 | Wide-angle imaging and neural network processing by the NNOD. a,** The setup for wide-angle imaging and neural network data collection on the optical platform. The images are projected onto a wide-angle display consisting of three OLED screens. The positions of the CMOS and metalens are adjusted to set the distance between them equals to metalens focal length. **b,** Sub-images directly captured by the three metalenses, each of which has its own clear imaging parts. **c,** Wide-angle image of letters pattern obtained

by stitching the clear image part of each sub-image, with a total FOV of 150°. **d,** Imaging result of the same letters pattern by a traditional hyperbolic phase metalens, the FOV of which is limited. **e,** Wide-angle imaging result of a complex image, with a FOV of 135°. **f,** Neural network processing. The three sub-images are initially processed by Module 1 for image quality enhancement. Subsequently, these images are stitched as a wide-angle image, which is then fed into Module 2 for intelligent object recognition and localization. **g,** The final wide-angle image and intelligent detection results after the neural network processing. The image highlights the recognition confidence of the target object and its $x$ and $y$ angle localization coordinates. **h,** Imaging result of the same complex scene by a traditional metalens for imaging FOV and quality comparison.

**Intelligent image processing and object detection**

Image quality plays a crucial role in the object detection performance[36-39], with higher-quality images typically yielding greater detection accuracy. While our metalens array has produced clear wide-angle images, there remains room for further improvement. Factors such as light transmission outside the virtual apertures, environmental stray light and intensity attenuation at large angles, can all contribute to the degradation of image quality. To address these issues, we proposed a comprehensive neural network (**Fig. 3f**), incorporating an adaptive image quality enhancement module (Module 1) and an intelligent image contents recognition module (Module 2).

Module 1 includes a multi-scale convolutional neural network (MSCNN) to extract hyperparameters (brightness, contrast, dynamic range, noise, etc.) of captured images, which would be sent to the subsequent multiple filters for image processing (**Supplementary Information Section 4**). The three sub-images were processed individually by Module 1 and then stitched together to form a complete high-quality wide-angle image. Before being fed into Module 2, a stitching artifact reduction operation is applied to further improve the image continuity (**Methods** and **Extended Data Fig. 4**). Module 2 is an improved network based on the YOLO network[40-43], incorporating a high-precision angular localization function. This module divides the input image into a grid and simultaneously predicts for each grid cell the bounding box coordinates, a confidence score indicating the likelihood of containing an object, and the associated class probabilities. The final prediction is made by selecting the bounding box with the highest confidence score for each detected object, and the angular localization is computed based on the predicted coordinates of the object's bounding box (**Supplementary Information Section 5**).

Through the neural network processing, not only a much higher quality wide-angle image is obtained, but also the objects in it can be intelligently detected (**Fig. 3g**). Compared with the traditional metalens imaging result (**Fig. 3h**), the detection information is greatly enhanced. To demonstrate the superior performances of the neural network, more comparison results of wide-angle images before and after neural network processing are also presented (**Extended Data Fig. 5**). Additionally, the car was put at different angular positions in the image, and then be detected by NNOD (**Methods** and **Extended Data Fig. 6**). The detection results indicate that even the car moves only 0.1°, the position change can be accurately identified, demonstrating the high-accuracy localization capability of NNOD.

Before conducting detection in real scenarios, we trained NNOD to possess the capability of recognizing various categories of objects. See **Supplementary Information Section 6** for detailed training process. To test the training effect, wide-angle images containing various objects (common categories include cars, person, truck, bus, UAV, bicycle, etc.) were projected on the wide-angle display for detection. The intelligent detection results demonstrate that the NNOD achieves a high recognition accuracy for all targets, averagely exceeding 96% (**Extended Data Fig. 7**). See **Supplementary Video S1** for wide-angle detection of dynamic various objects.

**High-mobility detection based on MAV integration**

We integrated the NNOD into a MAV (Fancinnov, FanciSwarm), enabling flexible wide-angle detection across various scenarios. The NNOD, along with other components such as the power supply module and signal transmission module, is mounted onto the MAV using a 3D-printed bracket (**Extended Data Fig. 8**). The images captured by the NNOD are transmitted through the interfaces of each module at high speed and ultimately sent wirelessly via Wi-Fi to a laptop (**Fig. 4a**). See **Supplementary Information Section 7** for detailed signal transmission process. The NNOD is integrated at the center of the top of 3D-printed bracket to prevent obstruction of the FOV (**Fig. 4b**). The zoom-in picture of NNOD and its architecture are shown in **Fig. 4c**, where a 3D-printed mask, a 550 nm optical filter, the metalens array and the CMOS sensor are compactly bonded together using optical clear adhesive (OCA). By adjusting the thickness of metalens array and CMOS sensor to match the metalens focal length, optimal imaging performances of far-field objects is achieved. **Supplementary Video S2** illustrates the capability of NNOD wirelessly transmit real-time captured images during flight.

At last, we realized the wide-angle imaging and intelligent detection of dynamic targets in a real street scene over a continuous period (**Supplementary Video S3**). Owning to the far-field imaging design of the metalens array, targets beyond ten meters can be effectively detected. A specific frame of the video is extracted to show the intelligent detection performance, including the various recognition categories and high accuracies of objects in the wide FOV scene (**Fig. 4d**). Notably, the recognition accuracy remains unaffected even when objects are positioned at large viewing angles, at the stitching seams of sub-images, or at areas where two objects overlap. The corresponding angular positions of these objects in this video frame are extracted respectively at $x$ and $y$ directions (**Fig. 4e** and **4f**). Each curve in these two figures corresponds to a target object in the wide-angle image (**Fig. 4d**), showing its continuous localization results from the moment it entered the wide-angle scene to the current frame. The numbers at the end of the curves indicate the specific localization results. The recognition and localization results of more video frames are also presented (**Extended Data Fig. 9**). It is worth noting that, due to the need to address alignment issues caused by image shake during the flight of MAV, as well as operations of image distortion correction, stitching and neural network processing, our current approach involves transmitting and storing the captured results for post-processing. With further improvements in the mechanical flight stability of MAVs and integration of all imaging processes into a unified software framework, our NNOD would enable real-time intelligent detection in broader future applications scenarios.

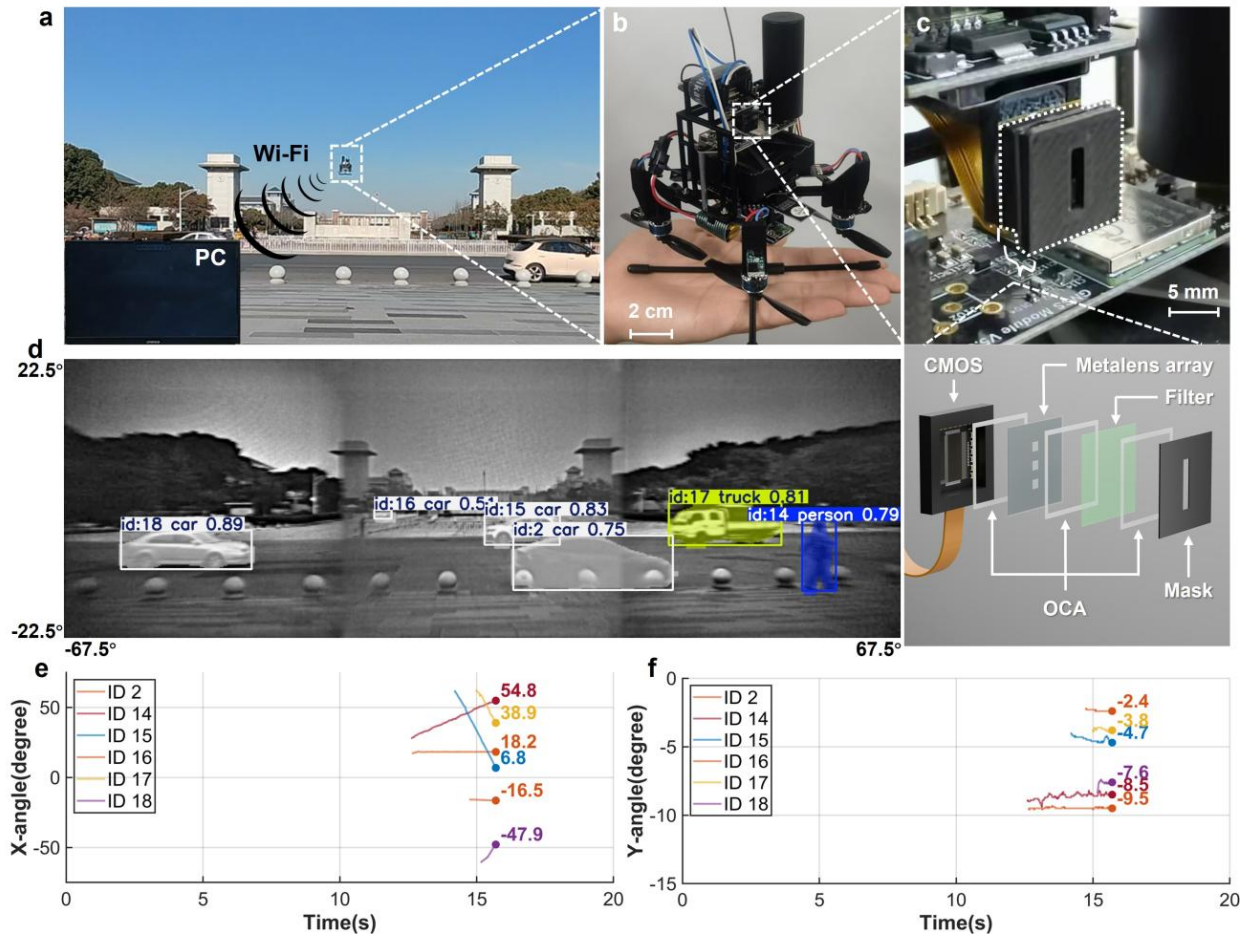

**Fig. 4 | High-mobility detection of real dynamic objects. a,** Schematic of high-mobility detection based on NNOD integrated with a MAV. Images captured by NNOD are wirelessly transmitted to a laptop through Wi-Fi signals. **b,** Photograph of the MAV integrated with NNOD. The MAV is 8 cm × 8 cm × 15.6 cm in size. The NNOD is integrated on top of the 3D-printed bracket. **c,** Zoom-in picture and architecture of the integrated NNOD. The mask, optical filter, metalens array and CMOS sensor are fixed together using optical optical clear adhesive (OCA). The distance between the metalens array and the CMOS is precisely controlled equaling to metalens focal length. **d,** One specific frame of the continuous detection results of a real wide-angle street scene. The text above each dynamic target represents the identification number, the recognized category and the recognition confidence/accuracy. The angular localization results of each target at **e,** the *x*-direction and **f,** the *y*-direction are also presented. The horizontal axis represents the entire time

of the continuous detection. Each curve corresponds to a target object in Fig. 4d, showing its continuous localization results from the moment it entered the wide-angle scene to the current frame. The numbers at the end of the curves are the localization results.

In conclusion, we developed an ultra-compact neural nanophotonic object detector (NNOD) with wide-angle imaging, intelligent recognition and localization capabilities. The system was integrated into a miniature unmanned aerial vehicle (MAV), demonstrating advantages in high-mobility detection and achieving high-performance operation across diverse real-world scenarios. The core component of the NNOD is a metalens array composed of three metalenses, achieving an imaging FOV exceeding 135°. Complemented by a comprehensive neural network, the NNOD delivers an average recognition accuracy of various objects over 80% and angular localization precision of 0.1°. The detailed performances comparison of our work and other object detection works is shown in **Supplementary Table S1**. Notably, the extension of 1D horizontal FOV presented in our work can accommodate most real-world applications on ground, such as wide-angle landscape photography, traffic intersection monitoring and autonomous driving detection. In addition, compared to refractive lenses with the comparable sizes, our metalens array achieves much wider FOV (**Extended Data Fig. 10**). Conventional wide-angle imaging typically requires complex fisheye lenses composed of multiple bulky refractive elements. In contrast, our compact wide-angle metalens array eliminates this limitation, enhancing the portability and integration of detection systems. Our work presents significant potentials for advancing the future developments of intelligent positioning, smart transportation, environmental monitoring etc. in the emerging low-altitude economy and the next-generation mobile communication technologies.

## Methods

### Ray tracing optimization

The ray tracing optimization process was conducted in Ansys Zemax OpticStudio software, aiming to obtain the phase profile of the center-FOV metalens. The optimized phase enables the metalens to achieve an imaging FOV of -22.5° to 22.5°. Due to the symmetrical design, we only need to analyze the light rays with incident angles within 0° to 22.5°. This angle range is divided into 4.5° intervals, resulting in light rays with 6 kinds of incident angles for analysis. Since the rays corresponding to each incident angle must be designed to exit through filter apertures located at different positions, a multi-configuration design strategy in Zemax is required, with each angle's rays assigned to a distinct configuration. In optimization, the objective is to minimize the average size of the focal spots for rays at the 6 incident angles on the focal plane, which could be expressed as:

$$[\phi(r), p_1, \ldots, p_6] = \arg\min_{\phi(r), p_1, \ldots, p_6} \sum_{i=1}^{6} S_i(\phi(r), \theta_i), \quad (1)$$

where, $\phi(r)$ is the metalens phase profile, $p_i$ is the distance between the metalens and the $i$-th filtering aperture, $\theta_i$ is the $i$-th incident angle, $S_i$ is the focal spot size of the $i$-th incident light. Although only 6 independent angles were analyzed to obtain the optimized phase, it was sufficient to achieve high-quality focusing for any angle within the continuous range.

In Zemax, the metalens is modeled as a "Binary2" surface type, which can be designed as a pure phase plane with no thickness. The phase profile was defined as an even order polynomial of the radial coordinate $r$ as

$$\phi(r) = \sum_{i=1}^{5} a_i \cdot \left(\frac{r}{R}\right)^{2i}, \quad (2)$$

where $R$ is the radius of the metalens, and coefficients $a_i$ were optimized for minimize the focal spots sizes. A fifth-order polynomial is sufficiently accurate to represent any symmetrical phase. The positions of the filtering apertures were also set as parameters to be optimized, which indirectly determines the regions of the metalens illuminated by incident light at different angles. By constraining the operands "TTLT" and "TTGT", which represent the minimum and maximum distances between the filtering apertures and the metalens, and by adjusting the weights of these operands in the "Merit Function Editor" in Zemax, a complex optimization objective tailored to our design can be achieved. Subsequently, employing the global optimization search in Zemax enables the acquisition of the optimal phase profile for the center-FOV metalens. The final optimized coefficients of the phase polynomial are shown in Table 1.

Table 1: Optimized coefficients of the phase polynomial

| $a_1$ | $a_2$ | $a_3$ | $a_4$ | $a_5$ |
|---|---|---|---|---|
| -888.0541 | 4.7298 | 1.8393 | -1.3519 | 0.0076 |

**FDTD simulation**

The transmission efficiency and electromagnetic field distribution of the metasurface unit cell were simulated with the finite-difference time-domain method via commercial software from Lumerical Inc., FDTD Solutions. Periodic boundary conditions were set in the *x* and *y* directions with a period of 275 nm to create an infinite periodic array of the unit cell as shown in **Extended Data Fig. 1a**. Perfect matching layers are employed along the z-axis to absorb the outgoing waves. The refractive indices of fused silica substrate and silicon nitride were from Handbook of Optical Constants of Solids (E.D. Palik)[44] and Optical Properties of Silicon Nitride (H. R. Philipp)[45] respectively.

**Fabrication of the metalenses**

Firstly, a 1000 nm-thick silicon nitride film was deposited on the $SiO_2$ substrate using Plasma enhanced chemical vapor deposition (PECVD). To enhance the imaging quality, a shading mask was then fabricated outside the metalenses area. After preprocessing, the substrates were loaded into the electron beam lithography system (Elionix ELS-F125) after sequentially coated with approximately 220 nm of resist film (AR-N7520) and about 50 nm of conductive adhesive (AR-PC 5090). After exposure, the conductive polymer was dissolved in water and resist was developed in a resist developer solution (ZX 238) for 40 s and fixer (DI water) for 60 s in sequence. Then the substrates were sequentially thermal evaporated a 10 nm thick chromium adhesion layer, a 50 nm thick aurum shading mask, and a 30 nm thick chromium layer as the protective coating for subsequent processing. After that, the samples were immersed in NMP solution for lift-out, leaving the shading mask adhered to the silicon nitride film. The following steps involve the fabrication of the metalens structure. A 225 nm PMMA A4 resist film and a 50 nm thick layer of conductive polymer (AR-PC 5090) were spin coated onto the sample in sequence and loaded into the electron beam lithography system (Elionix ELS-F125). After exposure, the conductive polymer was then dissolved in water and resist was developed in a resist developer solution (MIBK: IPA=1: 3) for 120 s and fixer (IPA) for 60 s in sequence. A 40 nm electron thermal evaporated chromium layer was used to reverse the generated pattern with a liftoff process, and was then used as a hard mask for dry etching the silicon layer. Finally, the sample is dry etched and is immersed in the stripping solution (ceric ammonium nitrate), the chromium layer is removed from the substrate, leaving only the desired metalens structure on the substrate. The whole fabrication process is shown in **Extended Data Fig. 1b**.

**Optical characterization**

The optical characterization includes two parts, the focusing performances characterization and the wide-angle imaging experiments. The setup of the focusing performances characterization is shown in **Extended Data Fig. 2a**. A 550 nm light filtered from a supercontinuum laser (NKT, SuperK COMPACT) is accurately controlled by an optical stage. The optical stage rotates along an arc with a radius of 50 cm, centered on the metalens array sample, and the emitted laser continuously faces the sample. When the laser rotates to different angles, the optical stage is adjusted accordingly to allow different metalenses to receive the light. The two side-FOV metalenses receive light at angles from -67.5° to -22.5° and from 22.5° to 67.5°, respectively, while the center-FOV metalens receives light at angles from -22.5° to 22.5°. After the laser beam is focused by the metalens array, the focal spots are captured by the CMOS image sensor (Imaging source, DMM 27UJ003-ML) for further analysis. The radial intensity distributions of the focal spots are extracted and plotted as the intensity distribution curves. The intensity distribution curves corresponding to different incident light are then normalized to the peak value of the 0° focal spot intensity.

The wide-angle images were projected onto a display composed of three organic light-emitting diode (OLED, CFORCE, CF015Next) screens. The screens were mounted on adjustable stands, allowing for continuous control of the entire viewing angle by adjusting the relative positions between each pair of screens. The imaging device composed of the metalens array and the CMOS, is mounted on a metal support rod to align with the center height of the screen and is connected to a desktop computer via a high-speed transmission line, enabling real-time transmission of the captured images.

**Imaging distortion correction**

The key step in distortion correction is establishing the mapping relationship between object points and image points. By applying the inverse transformation of this mapping, distortion can be corrected. Although the mapping is initially determined during the design of the metalens array using ray tracing methods, errors in sample fabrication and testing prevent its direct application. Traditional approaches, such as the Brown-Conrady model[46], express the object point coordinates as even-degree polynomial functions of the image point coordinates, with the polynomial coefficients determined through sampling and fitting to establish the mapping relationship. However, these traditional methods are primarily effective for addressing center-symmetric distortion types, such as pincushion or barrel distortion. As a result, for the side-FOV metalenses in our lens array, which experience asymmetric distortion, traditional methods are not sufficient. To address this, we define the mapping relationship as a polynomial that includes both odd- and even-degree terms, shown as

$$\begin{aligned}
x = f_a(x_d, y_d) &= a_1 + a_2 \cdot x_d + a_3 \cdot y_d + a_4 \cdot x_d^2 + a_5 \cdot x_d \cdot y_d + a_6 \cdot y_d^2 \\
&+ a_7 \cdot x_d^3 + a_8 \cdot x_d^2 \cdot y_d + a_9 \cdot x_d \cdot y_d^2 + a_{10} \cdot y_d^3 \\
y = f_b(x_d, y_d) &= b_1 + b_2 \cdot x_d + b_3 \cdot y_d + b_4 \cdot x_d^2 + b_5 \cdot x_d \cdot y_d + b_6 \cdot y_d^2 \\
&+ b_7 \cdot x_d^3 + b_8 \cdot x_d^2 \cdot y_d + b_9 \cdot x_d \cdot y_d^2 + b_{10} \cdot y_d^3
\end{aligned} \quad (3)$$

where, $x$ and $y$ refer to the object points coordinates, $x_d$ and $y_d$ refer to the image points coordinates, $a_i$ and $b_i$ are the coefficients to be determined. For both the center-FOV and side-FOV metalenses, 9×9 uniform 2D grid of points were selected, respectively and then be imaged by the metalenses. Through the fitting of these object points and image points, the polynomial coefficients can be determined. Through this polynomial, the distorted coordinates ($x_d$, $y_d$) can be substituted to calculate the undistorted coordinates ($x$, $y$), thereby achieving distortion correction (**Extended Data Fig. 3a**).

Another important issue is that if the ($x_d$, $y_d$) coordinates on the distorted image are traversed to calculate the corresponding ($x$, $y$) coordinates, it may lead to many ($x$, $y$) coordinates that not integer values. To address this issue, we traverse the integer pixel coordinates ($x$, $y$) to find the corresponding ($x_d$, $y_d$), and then use a quadratic interpolation method to determine the intensity of pixel ($x$, $y$) based on the intensities of the four nearest integer coordinates around ($x_d$, $y_d$). According to **equation (3)**, the corresponding coordinates ($x_d$, $y_d$) for each ($x$, $y$) are firstly determined, which are mostly non-integer pixels. The four integer points closest to ($x_d$, $y_d$) are labeled as ($x_{floor}$, $y_{floor}$), ($x_{floor}$, $y_{ceil}$), ($x_{ceil}$, $y_{floor}$), and ($x_{ceil}$, $y_{ceil}$), respectively. Then the intensity of pixel ($x$, $y$) would be calculated as

$$\begin{aligned}
Q(x, y) &= (y_{ceil} - y_d) \cdot \left[ Q_{11} \cdot (x_{ceil} - x_d) + Q_{21} \cdot (x_d - x_{floor}) \right] \\
&+ (y_d - y_{floor}) \cdot \left[ Q_{12} \cdot (x_{ceil} - x_d) + Q_{22} \cdot (x_d - x_{floor}) \right]
\end{aligned} \quad (4)$$

where, $Q_{11}$, $Q_{12}$, $Q_{21}$ and $Q_{22}$ represent the intensity of pixels ($x_{floor}$, $y_{floor}$), ($x_{ceil}$, $y_{floor}$), ($x_{floor}$, $y_{ceil}$) and ($x_{ceil}$, $y_{ceil}$), respectively (**Extended Data Fig. 3b**). After sequentially determining the intensity of all the pixels ($x$, $y$), the complete distortion-corrected image would be obtained. We then imaged a chessboard pattern using the metalens array and applied the distortion correction method. The comparison between the distorted and corrected image of a chess board (**Extended Data Fig. 3c and 3d**) highlights the high-quality distortion correction effectiveness achieved by our approach.

**Mitigating stitching artifacts**

The stitching seam is at the imaging edge of each metalens, where the greater coma aberration and significant intensity attenuation occurs. Therefore, to achieve a smooth transition between the two images, the intensity values of pixels within a certain range on both sides of the seam should be adjusted. Assuming the boundaries on both sides of the seam that require intensity adjustment are $L$ and $R$, with intensity values $I_L$ and $I_R$, respectively (**Extended Data Fig. 4a**). For any point $M$ within this range, its original intensity is $I_M$, and the adjusted intensity value is $I'_M$. The value $I'_M$ is obtained by multiplying $I_M$, $I_L$, and $I_R$ with their respective weight coefficients $W_M$, $W_L$ and $W_R$, expressed as $I'_M = W_M I_M + W_L I_L + W_R I_R$. The three coefficients are related to the distance of point M to the seam and the distances from $M$ to the two boundaries $L$ and $R$, which are respectively expressed as,

$$W_M = a \cdot |x_M - x_s|, \quad W_L = (1 - a \cdot |x_M - x_s|) \cdot \frac{x_R - x_M}{x_R - x_L}, \quad W_R = (1 - a \cdot |x_M - x_s|) \cdot \frac{x_M - x_L}{x_R - x_L}, \quad (5)$$

where, $x_M$, $x_L$, $x_R$ and $x_s$ are the $x$ coordinates of the point $M$, the left boundary, the right boundary and the seam respectively, and $a$ is an adjustable parameter that can be tuned according to different images. Notably, the configuration of **equation (5)** ensures that points closer to the seam have a smaller proportion of their original intensity, while those nearer to the boundaries have larger proportion, which preserves the continuity of intensity across the region from $L$ to $R$. **Extended Data Fig. 4b-d** present the comparison of the original image and the wide-angle imaging results before and after mitigating stitching artifacts. **Extended Data Fig. 4e** shows the comparison of intensity curve distributions within 20 pixels on both sides of the seam at different positions. The processed intensity distribution curve (blue) is noticeably smoother than the unprocessed intensity distribution curve (red) and aligns more closely with the intensity distribution curve in the original image (yellow).

**High-precision angular positioning scheme**

To ensure the positioning accuracy, it is necessary to first calibrate the positions of the objects in wide-angle imaging. Although distortion recovery algorithms have been employed, there still has slight deviation in angle calculation. To mitigate this deviation, we implemented a compensation algorithm. First, the target object car is placed at 20 different positions on the original image, and the wide-angle imaging results are captured and processed to remove distortion. Then, both the original images and the processed imaging results are subjected to the target recognition algorithm, producing 20 pairs of coordinates. The coordinates identified in the original images can be considered as the ground truth, while those obtained from the processed imaging results will exhibit certain deviations. These deviation values are then interpolated to create an error compensation curve across the full viewing angle range. Through this curve, the compensation angle for the coordinate position at any given angle can be obtained. By adding this compensation angle, accurate angular localization results can be achieved. The method for 2D compensation is like 1D compensation, with the key difference being that sampling is performed on a 2D plane, resulting in a fitted angle compensation surface.

After finishing the angular position calibration, we first tested the positioning capability by randomly put the object car at four different angular positions (-55.6°, -12.3°), (-9.9°, -8.4°), (9.1°, -8.6°) and (38°, -11.4°), respectively (**Extended Data Fig. 6a**). The two coordinates are angular position at *x* and *y* direction, respectively. After the imaging, recognition and processing of NNOD, and adding the aforementioned angular compensation, the positioning results of the four cars are completely consistent with the set angle coordinates (**Extended Data Fig. 6b**). Secondly, to further demonstrate the high-accuracy of our positioning method, the object car was put at another four positions, with the angular positions at *x* are 0.1° different from the previously set positions, that are (-55.7°, -12.3°), (-10.0°, -8.4°), (9.0°, -8.6°) and (37.9°, -11.4°), respectively (**Extended Data Fig. 6c**). The detection results show that even if the angular position difference is 0.1°, which can be hardly distinguished by the naked eye, it can still be accurately identified by NNOD (**Extended Data Fig. 6d**).

## Data Availability

The data that support the findings of this study are provided in the Supplementary Information/ Source Data file. Source data are provided with this paper.

## Code Availability

Source codes for the metalens array wide-angle imaging processing and object recognition and localization are available at: https://doi.org/10.5281/zenodo.****** (ref. 47). Additional codes are available upon request from the corresponding author Ji Chen (jichen@seu.edu.cn).

## Acknowledgements (optional)

This work is financially supported by National Key R&D Program of China (2022YFA1404301), National Natural Science Foundation of China (Nos. 12104223, 62325504, 61960206005 and 61871111), Jiangsu Key R &D Program Project (No. BE2023011-2), Young Elite Scientists Sponsorship Program by CAST (No. 2022QNRC001), the Fundamental Research Funds for the

## Ethics declarations

**Competing interests:** All other authors declare they have no competing interests.

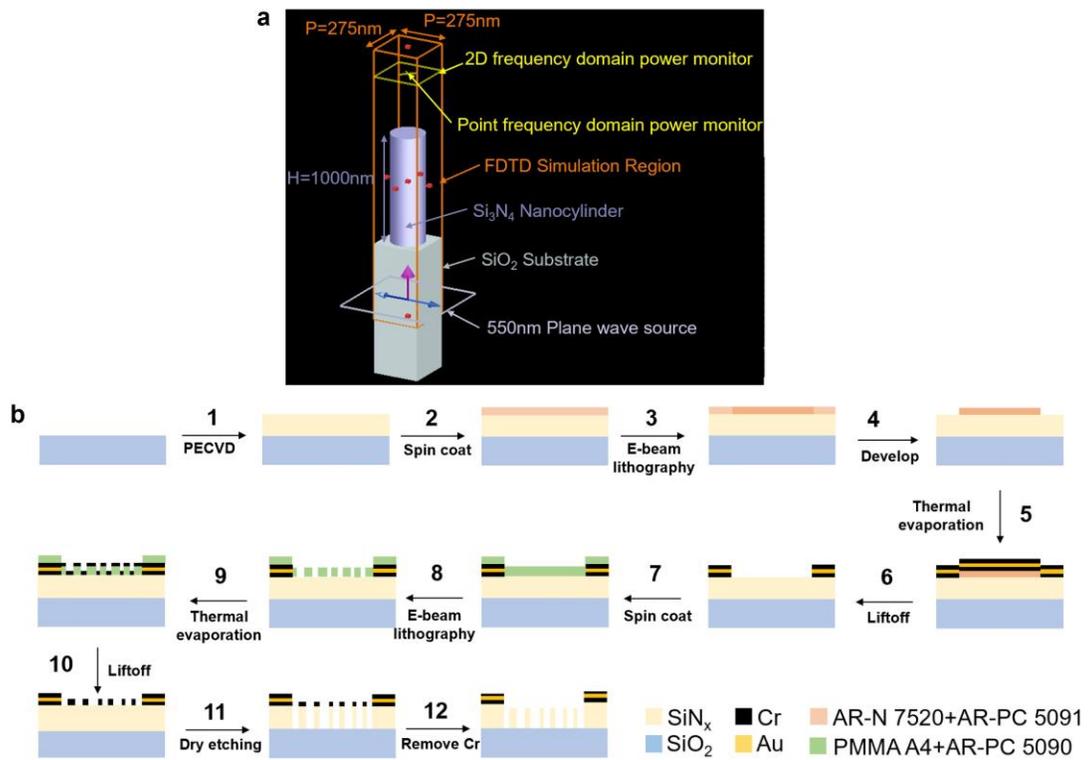

**Extended Data Fig. 1 | Design and fabrication of the metalens structure. a,** The simulation model in FDTD solutions. **b,** Schematic of the nano-fabrication process, detailed steps are shown in **Methods**.

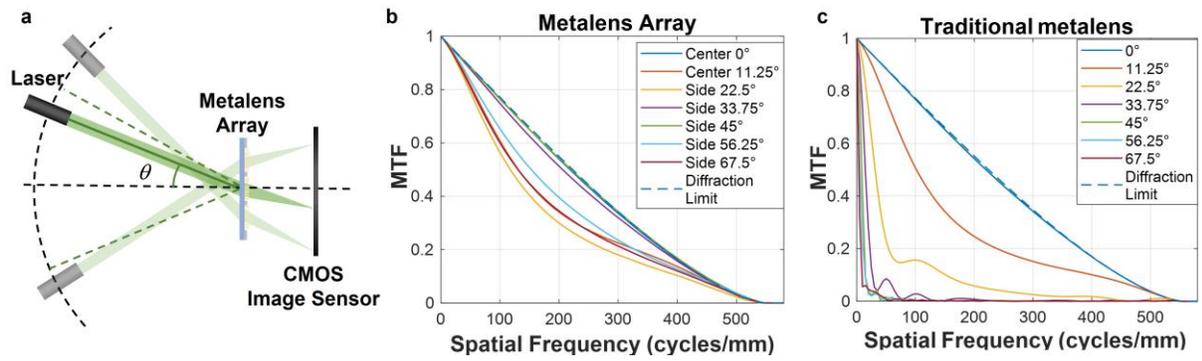

**Extended Data Fig. 2 | Characterization of wide-angle focusing performances of the metalens array. a,** The optical setup for laser incident at different angles. The MTFs of **b,** the wide-angle metalens array and **c,** a traditional metalens, corresponding to different incident angles. The MTF curves of the wide-angle metalens array are all close to the diffraction limit curve, demonstrating the good focusing performances.

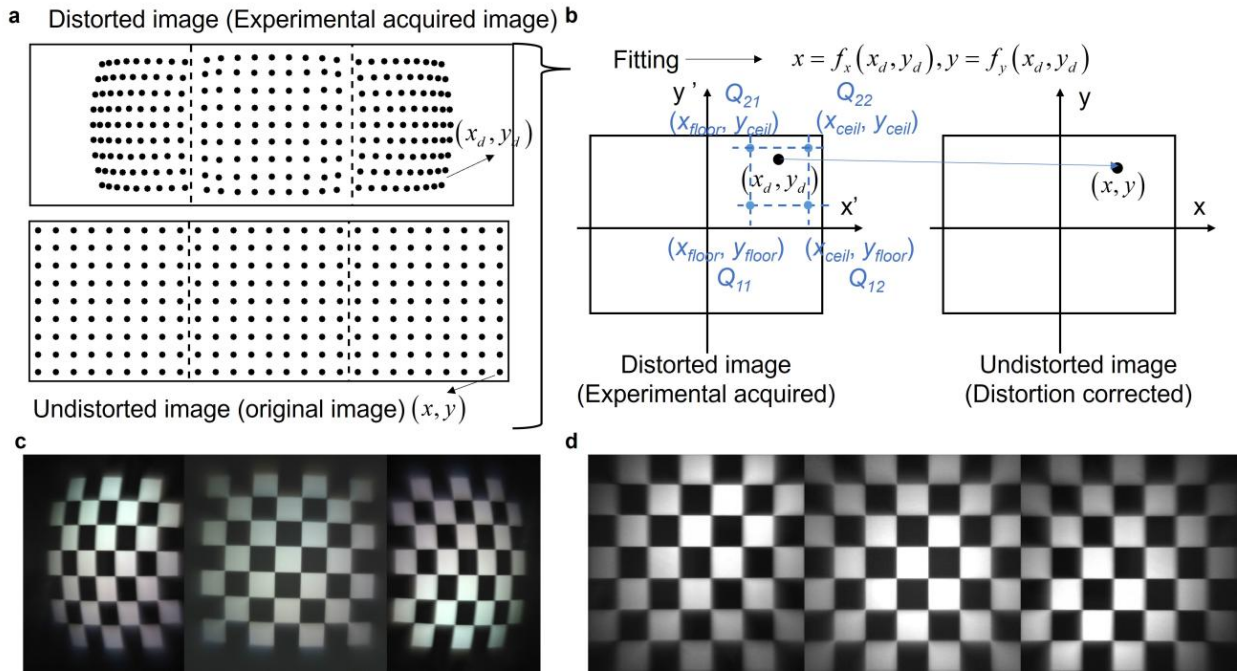

**Extended Data Fig. 3 | The distortion correction of metalens array wide-angle images. a,** The 2D sampling grids corresponding to the distorted image and the undistorted image, respectively. **b,** The sampling points correspondence between the distorted image and the undistorted image. The coordinate correspondence is determined by $x = f_x(x_d, y_d), y = f_y(x_d, y_d)$. The intensity of point $(x, y)$ is calculated by the average of the intensities of the four nearest integer coordinate points of $(x_d, y_d)$. Details are shown in **Methods**. **c,** The three distorted sub-images of a chess board directly captured by the metalens array. **d,** The wide-angle image of the chess board after distortion correction, in which the black and white squares are aligned horizontally and vertically, with almost no distortion.

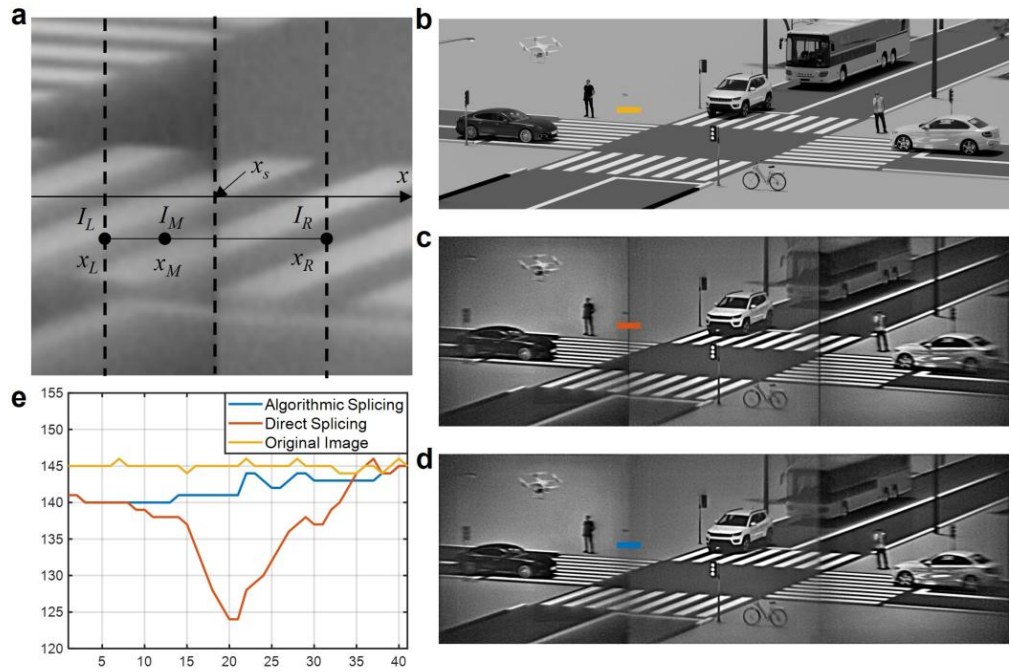

**Extended Data Fig. 4 | Mitigating stitching artifacts. a,** The area for intensity adjustment. The three dashed lines are the left boundary, the stitching seam and the right boundary, respectively. Points at these three dashed lines possess coordinates at *x* direction of $x_L$, $x_M$ and $x_R$, respectively, and intensities of $I_L$, $I_M$ and $I_R$, respectively. $x_M$ and $I_M$ are the coordinate and intensity of any point in this area, the intensity of which would be adjusted. The comparison of **b,** the original image, the wide-angle imaging results **c,** before and **d,** after stitching artifacts mitigation. **e,** The intensities of points within 20 pixels at both sides of the stitching seam. The yellow, red and blue curves are corresponding to the short lines shown in b-d, respectively. Results indicate that after processing, the intensity distribution is much smoother.

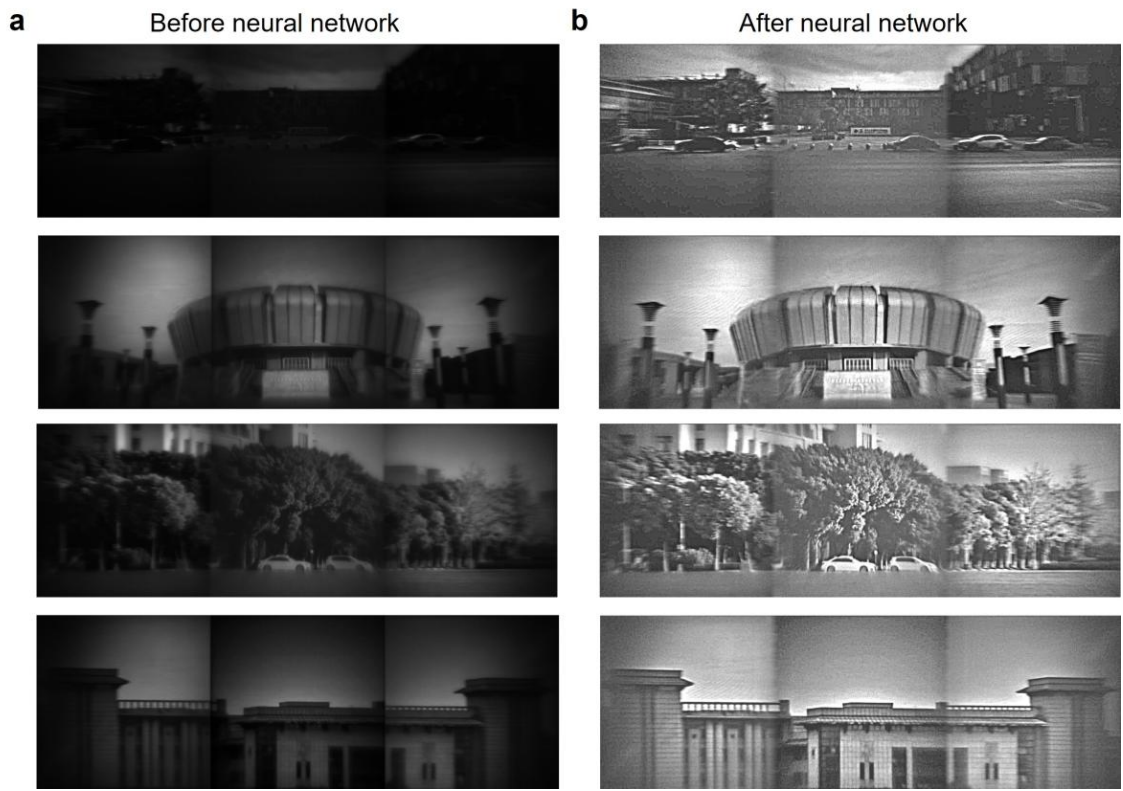

**Extended Data Fig. 5 | Comparisons of neural network processing.** Wide-angle imaging results **a,** before neural network processing and **b,** after neural network processing.

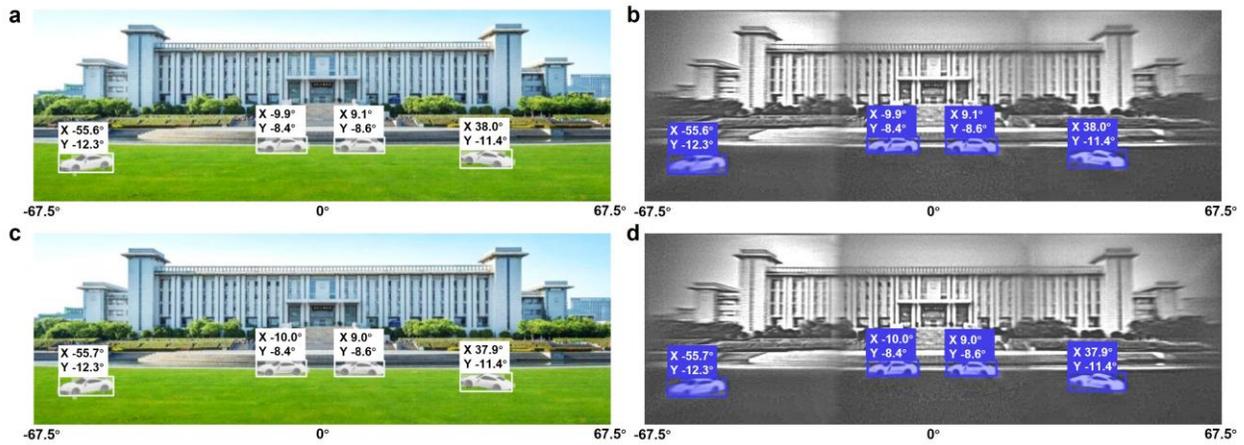

**Extended Data Fig. 6 | High-precision angular positioning based on NNOD. a,** Original wide-angle image where the target car is placed at four different angular positions (-55.6°, -12.3°), (-9.9°, -8.4°), (9.1°, -8.6°) and (38°, -11.4°), respectively, including both large and small angular positions. **b,** The NNOD detection results of Extended Data Fig. 6a, after angular compensation. The detected angular coordinates of the cars are completely consistent with those set in the original image. **c,** Original wide-angle image where the target car is placed at another four different angular positions (-55.7°, -12.3°), (-10.0°, -8.4°), (9.0°, -8.6°) and (37.9°, -11.4°), respectively, differing by 0.1° in *x*-direction from those in Extended Data Fig. 6a. **d,** The NNOD detection results of Extended Data Fig. 6c. Results show that even though the cars in the two original images differ by only 0.1°, they can still be accurately detected by the NNOD.

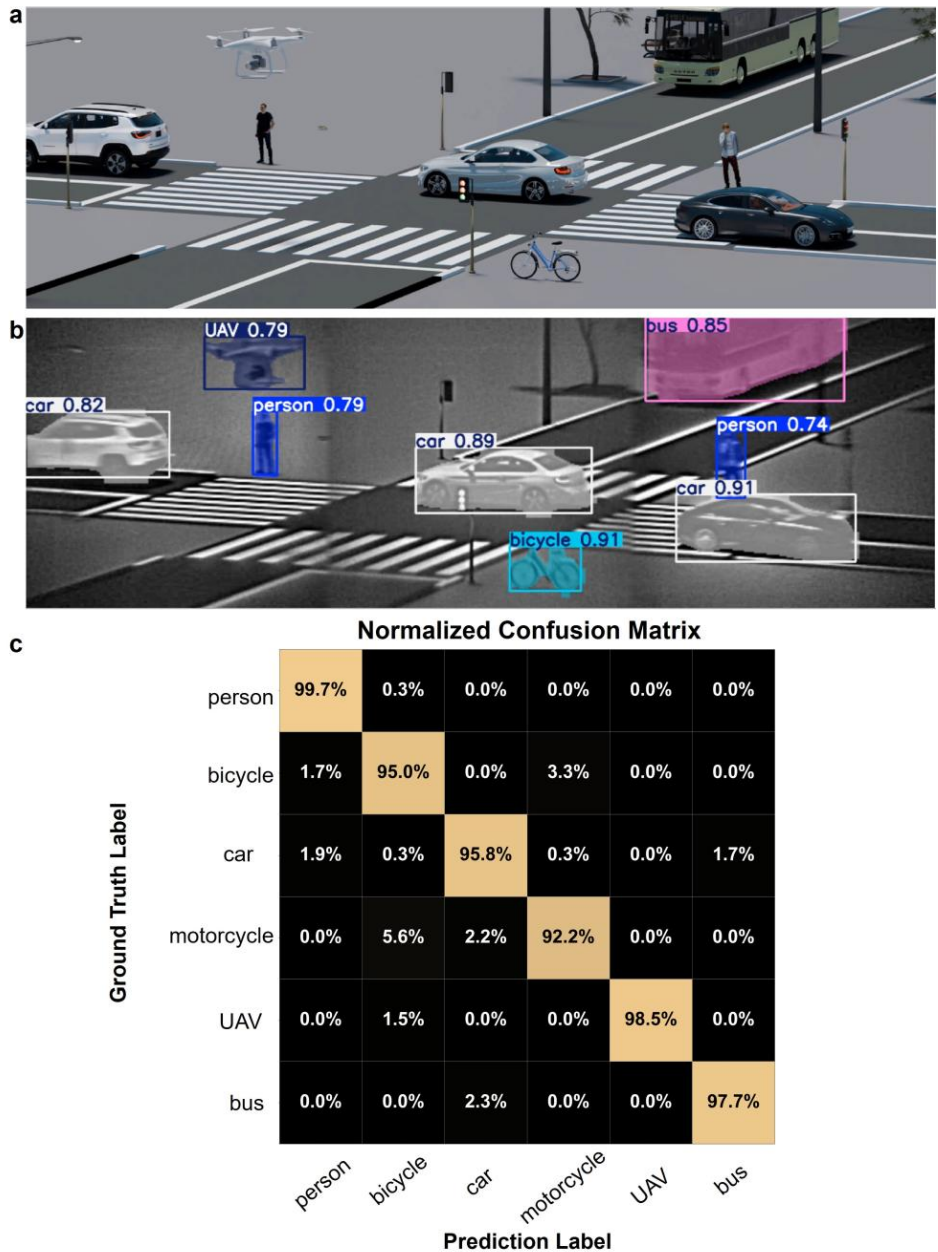

**Extended Data Fig. 7 | Recognition results of NNOD for various common object categories. a,** A custom-built wide-angle street scene containing multiple common categories of objects. The image was displayed on the wide-angle screen for NNOD detection. **b,** The detection results, including the recognized category and recognition confidence. **c,** The confusion matrix of the recognition results for several common categories in the validation dataset. Each category has 100 validation results, and the diagonal elements in the matrix represent the number of correct recognition results. It can be found that the average target recognition accuracy is higher than 96%.

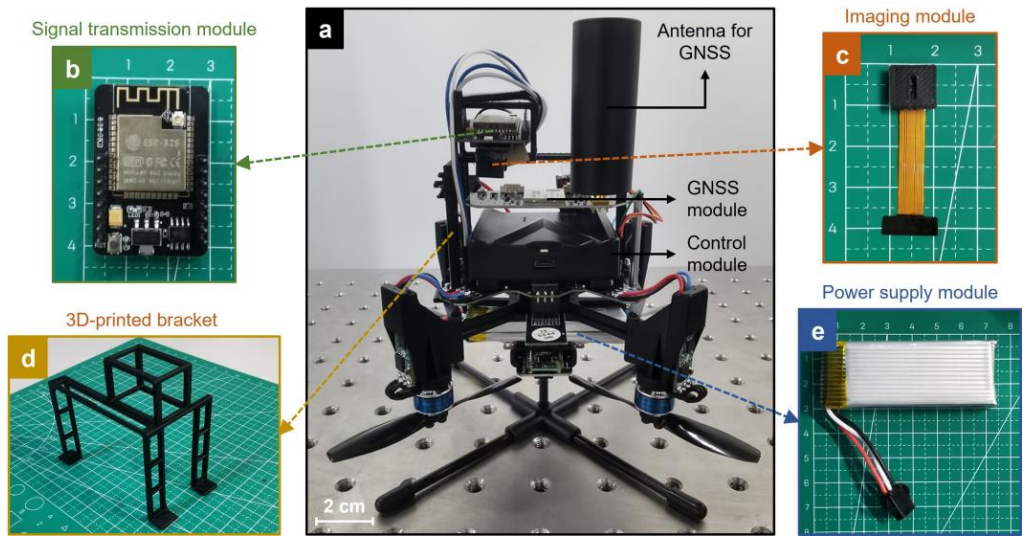

**Extended Data Fig. 8 | Integration details of the MAV. a,** Photograph of the MAV integrated with the NNOD. The GNSS module (global navigation satellite system) is applied to enhance the stability of MAV outdoor flight. **b,** Zoom-in picture of the signal transmission module (ESP32-S, dimensions: 4 cm × 3 cm × 0.5 cm, weight: 7 g). **c,** Zoom-in picture of imaging module (dimensions: 1 cm × 1 cm × 0.4 cm, weight: 0.5 g), the detailed components of which are shown in **Fig. 4c**. The signal transmission module receives information captured by the imaging module via a high-speed data cable and transmits it wirelessly to the receiving-end laptop. **d,** 3D-printed polymer support structure providing stabilization and secure mounting of integrated components. **e,** Zoom-in picture of the power supply module (7.4V lithium-ion battery with a capacity of 800 mAh), powering the imaging module, the signal transmission module and the MAV flight.

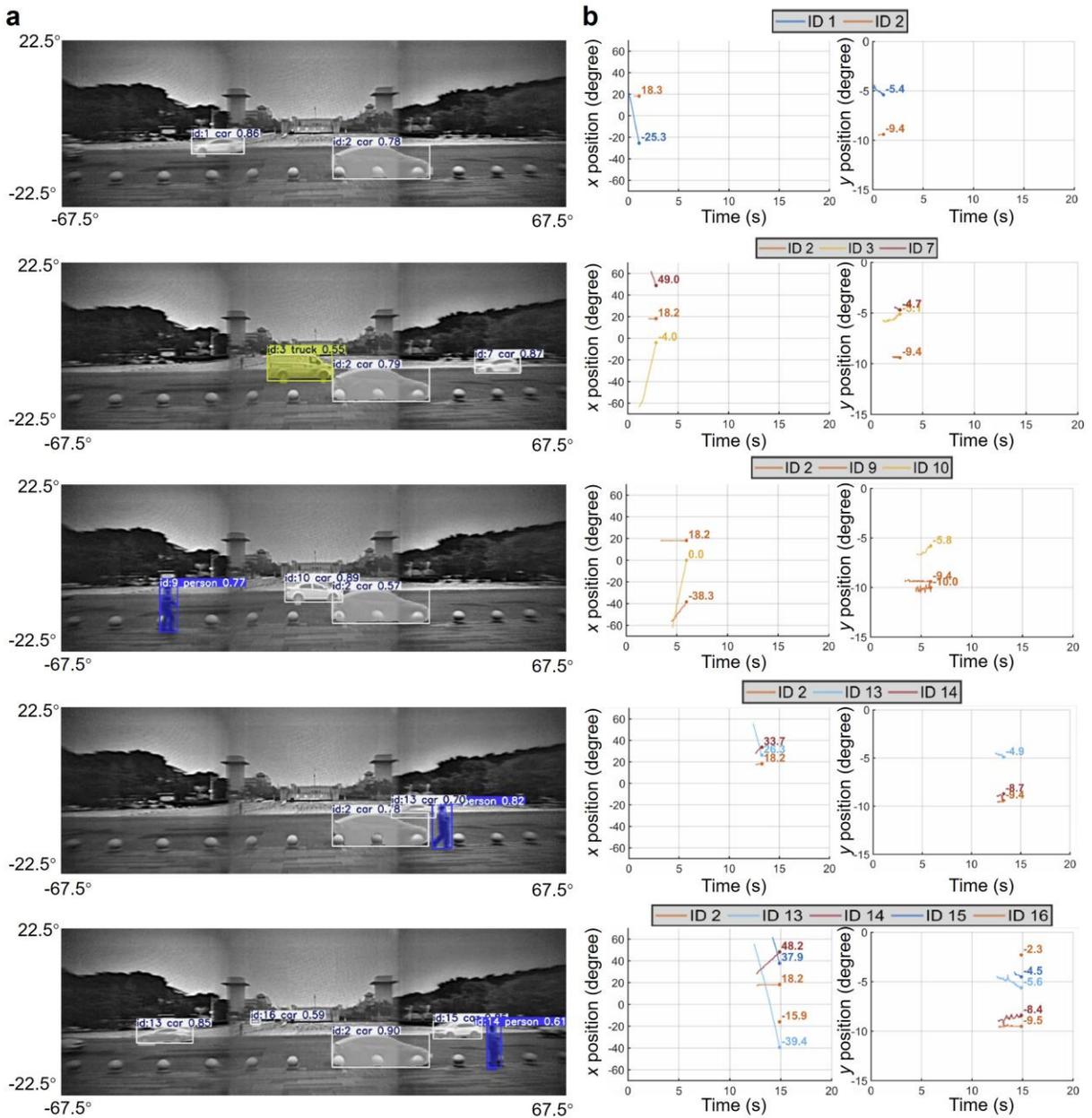

**Extended Data Fig. 9 | Multiple frames of Supplementary Video S3. a,** The targets recognition results of wide-angle scenes in different frame of Supplementary Video S3. The FOVs of these scenes are all 135° in *x* direction and 45° in *y* direction. **b,** The corresponding angular localization results in *x* and *y* directions. Each curve corresponds to a target recognized in the left-side wide-angle scene. The start and end points of the curve represent the time when the target enters the FOV and the time of the current frame, respectively. Therefore, each curve records the motion trajectory of the target within the wide FOV. The number at the end of the curve indicates the angular position of the target in the current frame.

Page **35** of **36**

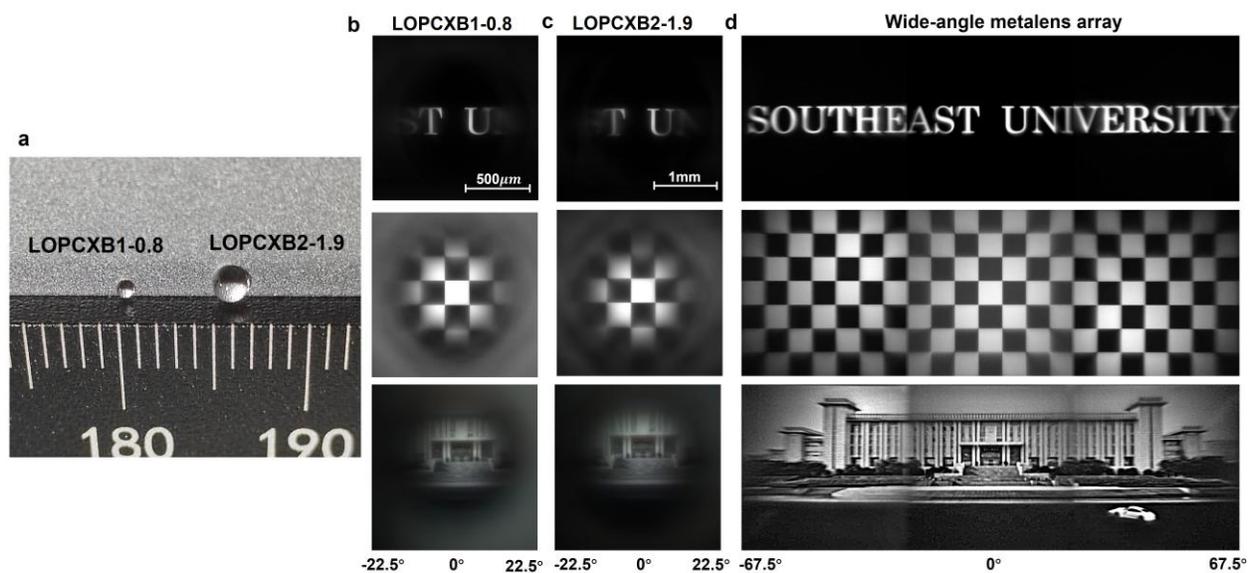

**Extended Data Fig. 10 | Imaging comparison between traditional refractive lenses and the wide-angle metalens array. a,** The two refractive lenses. **b-d**, Images captured by the two refractive lenses and our wide-angle metalens array. The results demonstrate that with the comparable sizes, our metalens array has much wider FOV.